# The Einstein specific heat model for finite systems

E. Boscheto*, M. de Souza**, and A. López-Castillo*◆

*Departamento de Química, Universidade Federal de São Carlos (UFSCar)

São Carlos, SP, Brazil

** IGCE, Unesp - Univ Estadual Paulista, Departamento de Física, 13506-900, Rio Claro, SP, Brazil

## Abstract

The theoretical model proposed by Einstein to describe the phononic specific heat of solids as a function of temperature consists the very first application of the concept of energy quantization to describe the physical properties of a real system. Its central assumption lies in the consideration of a total energy distribution among $N$ (in the thermodynamic limit $N \to \infty$) non-interacting oscillators vibrating at the same frequency ($\omega$). Nowadays, it is well-known that most materials behave differently at the nanoscale, having thus some cases physical properties with potential technological applications. Here, a version of the Einstein's model composed of a finite number of particles/oscillators is proposed. The main findings obtained in the frame of the present work are: (i) a qualitative description of the specific heat in the limit of low-temperatures for systems with nano-metric dimensions; (ii) the observation that the corresponding chemical potential function for finite solids becomes null at finite temperatures as observed in the Bose-Einstein condensation and; (iii) emergence of a first-order like phase transition driven by varying $N$.

** Present Address: Institute of Semiconductor and Solid State Physics, Johannes Kepler University - Linz, Austria.

◆Corresponding author e-mail: alcastil@ufscar.br

# 1 Introduction

The Einstein model [1] encompasses basic principles of Thermodynamics, Quantum and Statistical Mechanics to describe the specific heat of solids. Whereas this model does not exhibit "perfect" quantitative performance [2], it is the very first realistic solid state model to consider the effect of the crystal lattice vibrations on the thermodynamic properties, see e.g. [3, 4]. As a matter of fact, there are only a few realistic systems, whose multiplicities can be calculated using elementary methods, see e.g. Ref. [5] and references cited therein. Essentially, the model proposed by Einstein in 1907 [1] to describe the thermal properties of a simple crystalline solid, treating the solid as an array of atoms consisting of independent three-dimensional harmonic oscillators, is still of great interest [3, 4].

The aim of this work is to extend the model proposed by Einstein for the case of finite number of harmonic oscillators. To this end, mathematical functions describing analogous thermodynamic properties for finite solids such as the specific heat and the chemical potential were deduced. Although exhibiting, for $N \to \infty$, the thermodynamic behavior well-known from textbooks [6], such analogous functions have the advantage of being defined for any $N$, which allows one to explore how close to the thermodynamic behavior the properties of solids with low numbers of particles can be. In other words, the introduction of analogous thermodynamic functions extends the range of applications of thermodynamic, statistical and quantum mechanics, from macroscopic to microscopic scales.

Given the high interest in the physical properties of nano-materials, several models have been proposed to explain the peculiar properties of finite solids. Among them, it is worth mentioning reports taking into account surface effects [7-9], the form of the nanoparticles [10] and the presence of impurities [11]. Following Einstein's original framework, the model proposed here is based on how the total energy of the system of interest is distributed among their constituting oscillators, whose description is presented in the following. Initially, a closed-form function $S$, analogous to the thermodynamic entropy was obtained for systems with a finite number of particles (one-dimensional harmonic oscillators), see Section 2 for details. By taking the derivative of $S$ with respect to the energy $E$, one obtains the analogous thermodynamic temperature as a function of E for finite systems, i.e. $\tau(E, N) = (\partial S/\partial E)^{-1}$. Then, by numerically differentiating $E(\tau,N)$ with respect to $\tau$ the specific heat as a function of $N$, namely $\chi(N)$, for finite systems is deduced. Similarly, by taking the derivative of the entropy in relation to $N$, and employing a discrete form of the Leibniz integral rule (see Appendix B), the function $m$, analogous to the chemical potential for finite systems, was obtained.

Essentially, the main findings achieved in the present work are: (i) for different finite $N$, the $\chi$ vs. $\tau$ curves reproduce qualitatively the experimental behavior observed in the limit of low $\tau$; (ii) $m(\chi)$ converges quickly (upon increasing $N$) on the thermodynamic behavior of chemical potential (specific heat) even for low values of $N$; (iii) while the thermodynamic chemical potential is null only for $\tau \to 0$, the function $m$ for finite $N$ can be null even at finite temperatures ($\tau_0$), with $\frac{\tau_0 k_B}{\hbar \omega}$ being numerically equals to the inverse of the harmonic series, which converges slowly to zero as N grows; $k_B$ and $\hbar$ are, respectively, the Boltzmann and Planck constants, $\omega$ stands for the oscillators vibrational frequency. Furthermore, the present model can represent a bridge between finite systems and the thermodynamic limit by considering different values of N. Yet, the model can be employed in the description of the phononic specific heat of small clusters or nano-systems. Before starting with the theoretical discussion we stress that the physical quantities to be derived for finite systems are analogous to those well-known from text-books for the thermodynamic limit. The same holds true for the discussion about the phase transition like behavior to be discussed in Section 3.4. Yet, we stress that the present approach can be employed for one-, two- and three-dimensional systems as well, since that in the frame of Einstein's model the oscillators are uncoupled and thus the dimensionality and the arrangement of the system are no longer relevant.

## 2 Theory

The number of accessible eigen-states, $\Omega$, in a solid can be obtained by considering the sharing of M (= $E/\hbar\omega$ - N/2) energy quanta among N non-interacting oscillators. This approach is frequently discussed in the literature [6] and it is given by:

$$\Omega(E,N) = \frac{(M+N-1)!}{M!(N-1)!} = \frac{\left(\frac{E}{\hbar\omega}+\frac{N}{2}-1\right)!}{\left(\frac{E}{\hbar\omega}-\frac{N}{2}\right)!(N-1)!}, \quad (1)$$

For $N \geq 2$, $\ln(\Omega)$ can be reorganized exactly as (see Appendix A for generic M = F(E)):

$$\ln(\Omega(E,N)) = \sum_{i=1}^{N-1} \ln\left(\frac{\frac{E}{\hbar\omega}-\frac{N}{2}+i}{i}\right). \quad (2)$$

The analogous to the energy and entropy per particle functions in the thermodynamic limit are readily obtained from the definitions $u \equiv E/N$ and $s \equiv S/N$, respectively. By considering $S = k_B \ln(\Omega)$, it turns out that:

$$s \equiv S/N = \frac{k_B}{N} \sum_{i=1}^{N-1} \ln\left(\frac{\frac{u}{\hbar\omega}-\frac{1}{2}+\frac{i}{N}}{\frac{i}{N}}\right). \quad (3)$$

Using the relation $\partial s/\partial u = \partial S/\partial E = 1/\tau$, an expression can be found for the intensive function $\tau$, analogous to the thermodynamic temperature:

$$\frac{1}{\tau} = \frac{k_B}{\hbar\omega} \sum_{i=1}^{N-1} \left(\frac{E}{\hbar\omega}-\frac{N}{2}+i\right)^{-1} = \frac{k_B}{N\hbar\omega} \sum_{i=1}^{N-1} \left(\frac{u}{\hbar\omega}-\frac{1}{2}+\frac{i}{N}\right)^{-1}. \quad (4)$$

The functions $E = E(\tau,N)$, or $u = u(\tau)$, cannot be obtained analytically by simple inversion of the Eq. (4) for N > 4 since it involves polynomials of fifth or higher degree. However, numerical techniques for root finding problem as, for instance, Newton-Raphson can be used to obtain these functions with arbitrary precision. s and u are given in Table1 for different values of N. It is worth mentioning that u cannot be less than $\frac{\hbar\omega}{2}$ since the zero point energy is the lowest value possible, namely the entropy goes to zero when $u = \frac{\hbar\omega}{2}$. The behavior of $u_N(\tau)$ for various values of N is depicted in Fig.1.

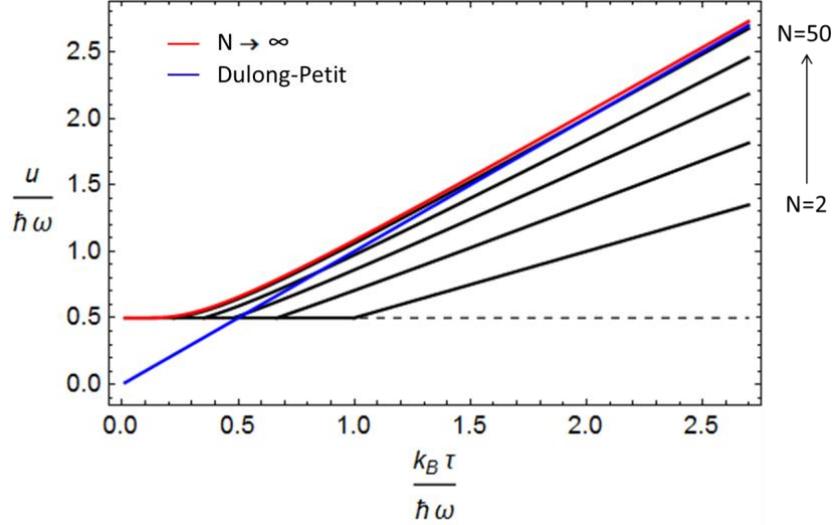

**Figure 1** – u vs τ for finite values of N (= 2, 3, 5, 10, 50 – black lines) and for N → ∞ (red line). The blue line indicates the classical energy u = $k_B T$ while the dashed line indicates the zero-point energy ($\frac{\hbar\omega}{2}$).

**Table 1** – The analogous energy (u) and entropy (s) per particle for some values of non-interacting one-dimensional oscillators N (and $N_{eff}$).

| N | $N_{eff}=N\pm N^{1/2}$ | s | u | u(τ→ 0) | s(u→ $\frac{\hbar\omega}{2}$) | u(τ→ ∞) |
|---|---|---|---|---|---|---|
| 2 | 2±1.41 | $\frac{1}{2}k_B \ln[\frac{2u}{\hbar\omega}]$ | $\frac{1}{2}k_B \tau$ | $\frac{\hbar\omega}{2}$ | 0 | $\frac{1}{2}k_B\tau$ |
| 3 | 3±1.73 | $\frac{1}{3}k_B \ln\frac{1}{2}[(\frac{3u}{\hbar\omega})^2 - (\frac{1}{2})^2]$ | $\frac{1}{3}\left(k_B\tau + \sqrt{\left(\frac{\hbar\omega}{2}\right)^2 + k_B^2\tau^2}\right)$ | $\frac{\hbar\omega}{2}$ | 0 | $\frac{2}{3}k_B\tau$ |
| k | k±$k^{1/2}$ | $\frac{k_B}{k}\sum_{i=1}^{k-1}\ln\left((\frac{u}{\hbar\omega} - \frac{1}{2} + \frac{i}{k})\frac{k}{i}\right)$ | Transcendental equation (not shown) | $\frac{\hbar\omega}{2}$ | 0 | $\frac{k-1}{k}k_B\tau$ |
| ∞ | ∞ | $k_B(\frac{u}{\hbar\omega} + \frac{1}{2})\ln[\frac{u}{\hbar\omega} + \frac{1}{2}] - k_B(\frac{u}{\hbar\omega} - \frac{1}{2})\ln[\frac{u}{\hbar\omega} - \frac{1}{2}]$ | $\frac{1}{2}\hbar\omega + \frac{\hbar\omega}{\exp\left(\frac{\hbar\omega}{k_B T}\right) - 1}$ | $\frac{\hbar\omega}{2}$ | 0 | $k_B T$ |

*Statistical fluctuation*

The effective number of particles, $N_{eff}$, can be modeled in any time by $N_{eff}$ = <N> + $(\sqrt{<N>})$ ran = <N> $(1 + \xi \, ran)$ = <N> I, where $\xi$ = <N>$^{-1/2}$ [12] is the relative amplitude order of the statistical fluctuations, I = N/<N> → 1 (for N → ∞) and -1 < ran < 1 is a number that follows a random sequence. The relative fluctuation of all properties varies as $\xi$ = <N>$^{-1/2}$. In this work, it was considered only the average value of effective number of particles to describe those properties, i.e., N ≡ <N>. $N_{eff}$ as a function of N is shown in Table 1.

## 3 Physical properties

### 3.1 Intensive properties

Since the systems of interest are finite, the extensivity of the thermodynamic potentials can be lost and as a result temperature and chemical potential are not true intensive functions as pointed out in the Introduction. $\tau$ as a function of N is given by Eq.(4). The N dependence of $\tau$ can be estimated using u = E/N = (M/N + 1/2)$\hbar\omega$ in Eq.(4) so that $\tau = \frac{\hbar\omega}{k_B}\left(\sum_{i=1}^{N-1}(M+i)^{-1}\right)^{-1}$. Hence, $\tau$ varies weakly with N as ~ (ln(N))$^{-1}$, i.e., $\tau$ can be considered nearly an intensive property. Another intensive property analogous to the chemical potential will be discussed below.

### 3.2 Specific heat

By employing the function u = u($\tau$) a simple differentiation leads to an expression for $\chi$, i.e. a function analogous to the thermodynamic molar specific heat, namely $\partial u/\partial \tau = \chi$. For low values of N, it is still possible to visualize analytically the values of the functions s, u and $\chi$. However, for N > 4 these values can be identified only by numerical techniques as Newton-Raphson which allows, inductively, for detecting general expressions, see Table 2. The molar specific heat becomes null when u becomes constant, i.e., u = $\frac{\hbar\omega}{2}$.

**Table 2** – Specific heat ($\chi$) and the analogous to the chemical potential for finite systems (m) for some values of N.

| N | $N_{eff}$=N±N$^{1/2}$ | $\chi$ | $\chi(\tau \to 0)$ | $\chi(\tau \to \infty)$ | m* | $\tau_0$ (m*=0) |
|---|---|---|---|---|---|---|
| 2 | 2±1.41 | $\frac{1}{2}k_B$ | 0 | $\frac{1}{2}k_B$ | $\frac{1}{\beta}\ln[\beta\hbar\omega]$ | $\frac{\hbar\omega}{k_B}$ |
| 3 | 3±1.73 | $\frac{1}{3}k_B\{1+\frac{k_B\tau}{\sqrt{\left(\frac{\hbar\omega}{2}\right)^2+k_B^2\tau^2}}\}$ | 0 | $\frac{2}{3}k_B$ | $-\frac{1}{\beta}\ln\left[\frac{3u_{N=3}}{2\hbar\omega}+\frac{1}{4}\right]$ | $\frac{2}{3}\frac{\hbar\omega}{k_B}$ |
| k | k±k$^{1/2}$ | *Transcendental equation (not shown)* | 0 | $\frac{k-1}{k}k_B$ | $-\frac{1}{\beta}\ln\left[\frac{k}{k-1}\frac{u_{N=k}}{\hbar\omega}+\frac{k-2}{2(k-1)}\right]$ | $\left(\sum_{i=1}^{k-1}i^{-1}\right)^{-1}\frac{\hbar\omega}{k_B}$ |
| ∞ | ∞ | $k_B\left(\frac{\hbar\omega}{k_BT}\right)^2\frac{exp\left(\frac{\hbar\omega}{k_BT}\right)}{\left(exp\left(\frac{\hbar\omega}{k_BT}\right)-1\right)^2}$ | 0 | $k_B$ | $\frac{1}{\beta}\ln\left[1-exp\left(\frac{-\hbar\omega}{k_BT}\right)\right]$ | 0 |

The general behavior of $\chi$ versus the reduced temperature, $\theta$ ($= \frac{\tau}{\theta_E}$, with the Einstein temperature defined as $\theta_E \equiv \frac{\hbar\omega}{k_B}$), is shown in Fig. 2 for several N´s. Note that if the energy of the solid is distributed only between two oscillators one has $\chi = k_B/2$, i.e., in this particular case $\chi$ would be independent of $\theta$. There is an increasing availability of oscillators to receive energy quanta $\chi(\theta)$ approaches to the thermodynamic behavior, $N \rightarrow \infty$ (red curve), where the specific heat assumes the smallest (low temperature) and the largest (high temperature) possible values, as shown in Table 2. It is worth mentioning here the similarity between the curves for N = 50 and $N \rightarrow \infty$, which suggests that the thermodynamic behavior can be sketchily exhibited already by small systems with a number of oscillators of a few tens.

Interestingly enough, upon increasing N the set of $\chi(\theta)$ curves show a first-order phase transition like behavior with a maximum (cf. red solid circles in Fig. 2) centered at $\theta \sim 0.40$ as depicted in Fig. 2. Following detailed discussion presented in Ref. [4], for any "Einstein´s solid" $\theta_E$ (the Einstein temperature) can be estimated roughly by taking the maximum of the plot $\chi$ over temperature *versus* temperature. The present results indicate that for finite systems $\theta_E$ can be tuned upon varying N. For completeness, it is worth mentioning that for an ordinary non-interacting bosonic system the condensation temperature strongly depends on the number of bosons [13].

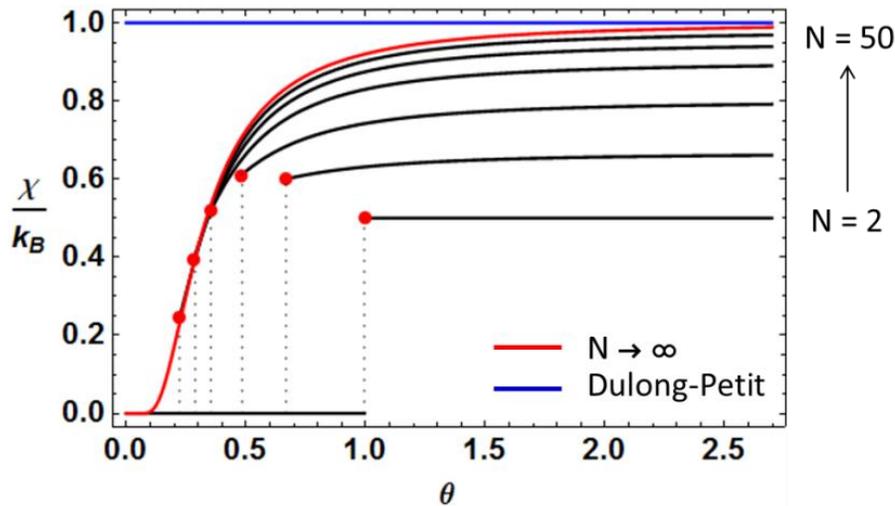

**Figure 2** – $\chi$ vs θ for finite values of N (= 2, 3, 5, 10, 20, 50 – black lines) and for $N \rightarrow \infty$ (red line). The blue line indicates the specific heat of Dulong-Petit, while the red points indicate the specific heat jump and entropy become null and the energy reaches to the zero point energy for different N.

As it will be discussed below in Section *3.3 Chemical Potential*, the jump to zero in the specific heat for finite N is closely related to the change of the system zero-point energy.

### 3.3 Chemical Potential

Essentially, the chemical potential gives the change of the free energy per particle when oscillators are added or removed from the system. In particular, the chemical potential is negative for an ensemble of harmonic oscillators and increases with decreasing of number of oscillators (*N*) as shown below. For a determined finite *N* the chemical potential becomes less negative as the temperature decreases and eventually reaches zero together with entropy at some critical value.

An expression for the analogous thermodynamic chemical potential, $m$ ($=-\tau\frac{\partial S}{\partial N}$), can be formulated by taking the derivative of *S* with respect to *N* (and by using the Leibniz integral rule [14] for discrete systems, see details in the Appendix B):

$$\frac{\partial S}{\partial N} = k_B \ln\left[\left(\frac{E}{\hbar\omega} + \frac{N}{2} - 1\right)\frac{1}{N-1}\right] - \frac{k_B}{2}\sum_{i=1}^{N-1}\left(\frac{E}{\hbar\omega} - \frac{N}{2} + i\right)^{-1} = -\frac{m}{\tau}. \quad (5)$$

By considering $\beta = \frac{1}{k_B \tau}$ and, from Eq. (4), $\sum_{i=1}^{N-1}\left(\frac{E}{\hbar\omega} - \frac{N}{2} + i\right)^{-1} = \frac{\hbar\omega}{k_B \tau}$, the Eq. (5) can be rewritten for any *N* as

$$\frac{E}{\hbar\omega} = (N-1)e^{-\beta\left(m - \frac{\hbar\omega}{2}\right)} - \frac{N}{2} + 1, \quad (6)$$

which for $N \to \infty$ reduces to:

$$\frac{u}{\hbar\omega} = e^{-\beta\left(m - \frac{\hbar\omega}{2}\right)} - \frac{1}{2}. \quad (7)$$

The Eq. (6) can also be written explicitly for *m* as a function of *E* and *N* as:

$$m = \frac{\hbar\omega}{2} - \frac{1}{\beta}\ln\left[\frac{\left(\frac{E}{\hbar\omega} + \frac{N}{2} - 1\right)}{N-1}\right], \quad (8)$$

with $m(\tau \to \infty) = -\frac{1}{\beta}\ln\left[\frac{k_B\tau}{\hbar\omega}\right]$ for any *N*.

The chemical potential is equal to the quantum vacuum zero point energy for $\tau \to 0$, i.e., $m(\tau \to 0) = \frac{\hbar\omega}{2}$ or $m^*(\tau \to 0) = \left(m - \frac{\hbar\omega}{2}\right) = 0$. In order to prove the latter statement, the following analysis is made:

$$\beta m^* = \beta\left(m - \frac{\hbar\omega}{2}\right) = -\ln\left[\frac{N}{N-1}\frac{u_N}{\hbar\omega} + \frac{N-2}{2(N-1)}\right], \quad (9)$$

note that *m\** is always negative.

If $u_{N\to\infty} = \frac{1}{2}\hbar\omega + \frac{\hbar\omega}{\exp\left(\frac{\hbar\omega}{k_BT}\right)-1}$ then $m^* = \frac{1}{\beta}\ln\left[1 - \exp\left(\frac{-\hbar\omega}{k_BT}\right)\right]$ [15] as shown in Table 2.

The intensivity of $m^*$ (or $\beta m^*$) can be analyzed by considering Eq.(9). The $\beta m^*$ quantity is virtually an intensive property since it varies as a logarithmic function of simple relations of $N$. Using $u = (M/N+1/2)\hbar\omega$ in Eq.(9) one obtains $\beta m^* = -\ln\left[\frac{N+M-1}{N-1}\right]$. Alternatively, if $u$ ($\tau \to \infty$) $= \frac{N-1}{N}k_B\tau$ in Eq.(9) one has $\beta m^* = -\ln\left[\frac{k_B\tau}{\hbar\omega} + \frac{N-2}{2(N-1)}\right]$, which varies as $\beta m^* \sim \ln[\ln(N)]$, i.e., $\beta m^*$ varies very slightly with $N$.

Note that $\beta m^*$ becomes null when $u_N = \frac{\hbar\omega}{2}$ (zero point energy) for any $N$ at temperature $\tau_0$. Such a behavior is analogous to the Bose-Einstein condensate temperature [16], i.e., below a certain temperature the chemical potential vanishes. Thus, considering the analogy with an ordinary Bose-Einstein condensation $\tau_0$ can be estimated as $\frac{\hbar\omega}{2k_B}$ by considering one of the four following equivalent situations:

a) $\lambda_{dB}$ (thermal de Broglie wavelength) $\sim <x>$ (average inter-particle distance), where $\lambda_{dB} = h/p = h/(2m<E>)^{1/2} = h/(2mk_B\tau_0)^{1/2}$ and $<V> = k<x>^2/2 = \frac{\hbar\omega}{2}$ (lowest energy);

b) $\Delta x \Delta p \sim h/2$, with $E = K + V$, where $<K> = <p>^2/2m$, $<V> = k<x>^2/2$, $k = m(\omega/2\pi)^2$, and $<V>/<K> = 1$ for the harmonic oscillator virial relation;

c) $\Delta E \Delta t \sim h/2$, with $\Delta E \sim k_B\tau_0$ (thermal energy) and $\Delta t \sim 2\pi/\omega$;

d) the classical curve, $u = k_B\tau$, touches the zero point energy, $u = \frac{\hbar\omega}{2}$ (see Fig.1) at $\tau_0 = \frac{\hbar\omega}{2k_B}$.

From Eq. (9), if $u_N = \left(\frac{\hbar\omega}{2} + \delta\right)$ then $\left(\frac{N}{N-1}\frac{u_N}{\hbar\omega} + \frac{N-2}{2(N-1)}\right) = \left(1 + \frac{N}{N-1}\frac{\delta}{\hbar\omega}\right) \equiv \zeta$. If $\delta \geq 0$ (for $N > 1$) then $\zeta \geq 1$ and $-\ln[\zeta] = \beta m^* \leq 0$. Since the zero point energy is the lowest value, then $\delta < 0$ is not allowed and consequently $\beta m^*$ cannot be positive.

Using $u_N = \frac{\hbar\omega}{2}$ in Eq. (4), it turns out that:

$$\tau_0 = \left(\sum_{i=1}^{N-1} i^{-1}\right)^{-1} \frac{\hbar\omega}{k_B}, \qquad (10)$$

where $\sum_{i=1}^{N-1} i^{-1}$ is the harmonic series, which diverges as $\ln(N)$ for $N \to \infty$ and in this limit $T_0 \to 0$. $\tau_0$ can also be given in terms of $\theta \equiv \frac{\tau k_B}{\hbar\omega}$, as follows:

$$\theta_0 = \left(\sum_{i=1}^{N-1} i^{-1}\right)^{-1}. \qquad (11)$$

Note that $\tau_0$ (or $\theta_0$) can also be obtained by using $u_N(\tau_0) = \frac{\hbar\omega}{2}$ in Eq(4) as shown below:

$$\frac{1}{\theta_0} = \frac{\hbar\omega}{k_B \tau_0} = \frac{1}{N}\sum_{i=1}^{N-1}\left(\frac{\frac{\hbar\omega}{2}}{\hbar\omega} - \frac{1}{2} + \frac{i}{N}\right)^{-1} = \sum_{i=1}^{N-1} i^{-1}.$$

The strictly negative behavior of the $\beta m^*$ function against the reduced temperature ($\theta$) for finite systems ($N$ = 2, 4, 10, and 50) is depicted in Fig. 3. For comparison, the thermodynamic limit (dashed curve) is also shown. It can be noted that in the frame of the approach considered here systems consisting of a few tens of particles (oscillators) are already in the thermodynamic regime at high temperatures, i.e., $\theta \gg \theta_0$.

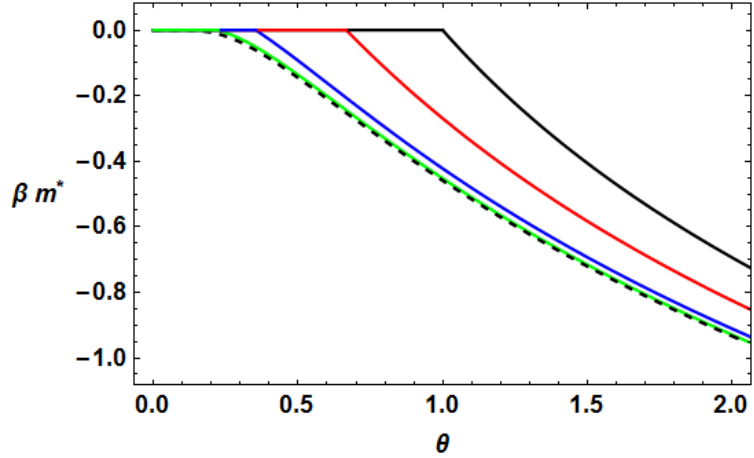

**Figure 3** – Normalized chemical potential (cf. Eq. (9)) as a function of θ for distinct number of oscillators N, N = 2 (black), 3 (red), 10 (blue), 50 (green), and $N \to \infty$ (dashed black line).

In our analysis, the assumption of $\tau_0$ as the Bose-Einstein condensate temperature is based on the following argument: when the chemical potential and entropy goes to zero at finite temperature the system spontaneously condensate upon decreasing temperature. It is well-known from classical textbooks that such a boson condensation cannot take place in fermions and classical particles, since for positive chemical potential (at low temperatures) a repulsive behavior appears once the system temperature is decreased. The aforementioned repulsion is related to the Pauli Exclusion Principle for fermions and to rigid barrier for classical particles, which prevents the condensation.

As can be inferred from Eq. (11), and directly seen in Fig. 4 (a), $\theta_0$ slowly decreases as a function of N proportionally to the inverse of the harmonic series. In terms of comparison, $\theta_0(N=10)$ = 0.3535 and $\theta_0(N = 10^{12})$ = 0.0355, i.e., the reduced Bose-Einstein condensate temperature shows a slight variation with N and it can still be relatively high for mesoscopic systems, i.e., $\theta_0 \approx (\ln(N))^{-1}$. The approximation $\left(\sum_{i=M}^{N} i^{-1}\right)^{-1} \approx (\ln(N) - \ln(M))^{-1}$ was considered for large values (>$10^6$) of N and M.

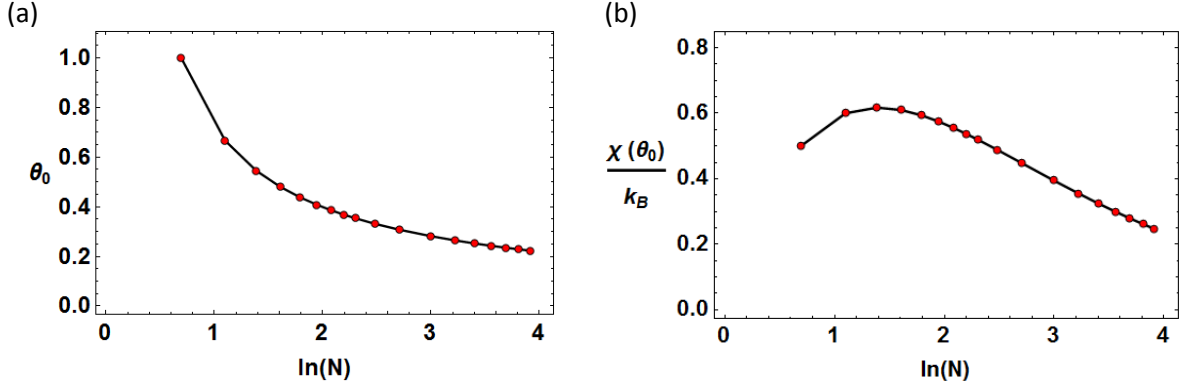

**Figure 4** – (a) The critical temperature, $\theta_0$, as a function of ln(N) (cf. Eq. (11)); (b) The analogous specific heat, $\chi$, evaluated at the analogous Bose-Einstein condensate temperature, $\theta$, as a function of ln(N). Solid lines are guide for the eyes. For details, see the main text.

The reduced critical temperature ($\theta_0$) is shown in Fig. 4 (a), where one can see the slow decreasing of $\theta_0$ as a function of $N$ (note the logarithm scale for $N$). The specific heat for the *analogous* to the Bose-Einstein condensate temperature for solids with different numbers of oscillators is shown in Fig. 4 (b). Similarly, it is possible to observe the slow convergence for null specific heat for increasing $N$.

For an ensemble of harmonic oscillators the critical temperature is null ($\theta_0 \to 0$) for the thermodynamic limit ($N \to \infty$). However, $\theta_0$ is finite for instance in the case of a boson gas trapped in a box. For the latter the eigen-values of the particles are $E_i = h^2 n_i^2/(8\, m\, L^2)$, where $L$ is the box length, being the zero point energy given by $E_{zp} = h^2/(8\, m\, L^2)$. $E_i$ around the critical temperature can be approximated by $E_i = E_{zp} + \delta$, with $\delta \ll 1$. Consequently, the energy quanta function $F(E) \approx N/2 + 4mL^2E/h^2$ or $f(u) \approx \tfrac{1}{2} + 4mL^2u/h^2$ since $M = \sum_{i=1}^{N} n_i = \sum_{i=1}^{N}\left(\frac{8mL^2 E_i}{h^2}\right)^{1/2}$. Finally, one can obtain $\theta_0 \approx \frac{1}{2}\left(\sum_{i=1}^{N-1}(i+N)^{-1}\right)^{-1}$ around the critical temperature, where $\theta \equiv \frac{8mL^2\tau k_B}{h^2}$. The critical temperature for $N \to \infty$ is $\theta_0 \approx \frac{2}{\ln 2}$ = 2.89, i.e., it is no null.

*3.4 Finite size phase transition*

Before starting to discuss the details of the quantities of interest in connection with a phase transition like behavior, we stress that first-order transitions have been intensively explored in the literature for finite size systems, see e.g. [17].

The $u_N(\tau)$ and $\beta m^*(\tau)$ show an equivalent finite size first-order transition like behavior at $\tau_0$, i.e., the first derivative $du_N/d\tau$ and $d\beta m^*/d\tau$ are discontinuous. For example, $u_{N=2}=\tfrac{1}{2}k_B\,\tau$ for $\tau > \tau_0$ and $u_N(\tau_0) = \tfrac{\hbar\omega}{2}$ for $\tau \leq \tau_0$, i.e., $du_{N=2}/d\tau = \tfrac{1}{2}k_B$ for $\tau > \tau_0$ and $du_N/d\tau = 0$ for $\tau \leq \tau_0$. Such discontinuous behavior of the first derivative appears for all $N$'s. Note that the energy ($u$) is described by a smoothed curve as a function of *T* only in the thermodynamic limit for classical and quantum models, cf. Fig.1. Hence, as well-known from textbooks, the Einstein solid model does not exhibit phase transition in the thermodynamic limit. However, such a behavior is not observed for finite systems. Instead, the energy (*u*) for finite systems is not a smoothed curve and consequently its first derivative, namely $du_N/d\tau$, is discontinuous. The heat capacity ($du_N/d\tau = \chi$), i.e. the first derivative of $u_N(\tau)$, shows for all *N* a discontinuity in $\tau = \tau_0$ as shown in Fig.2. These jumps appear at $\tau = \tau_0$, where the $u_N(\tau)$ curves are not smoothed.

Yet, *m\** shows similar first-order transition like behavior. $\beta m^*$ is given by Eq.(9) (Fig.3) for $\tau > \tau_0$ and $\beta m^* = 0$ for $\tau \leq \tau_0$, which implies in $d\beta m^*/d\tau \approx -\tau^{-1}$ for large $\tau$ and $d\beta m^*/d\tau = 0$ for $\tau \leq \tau_0$. That is, the derivative $d\beta m^*/d\tau$ shows a discontinuity as shown in Fig.5. The $d\beta m^*/d\tau$ is given by:

$$\frac{d\beta m^*}{d\tau} = -\frac{\chi}{u_N + \frac{N-2}{N}\frac{\hbar\omega}{2}}$$

with for $\tau \to \infty$ ($\theta \to \infty$)

$$\frac{d\beta m^*}{d\theta} \approx -\frac{1}{\theta + \frac{N-2}{2(N-1)}} \approx -\theta^{-1}$$

Note that, the $d\beta m^*/d\tau$ is proportional to $\chi$. In particular, this derivative shows simple behavior for large $\tau$.

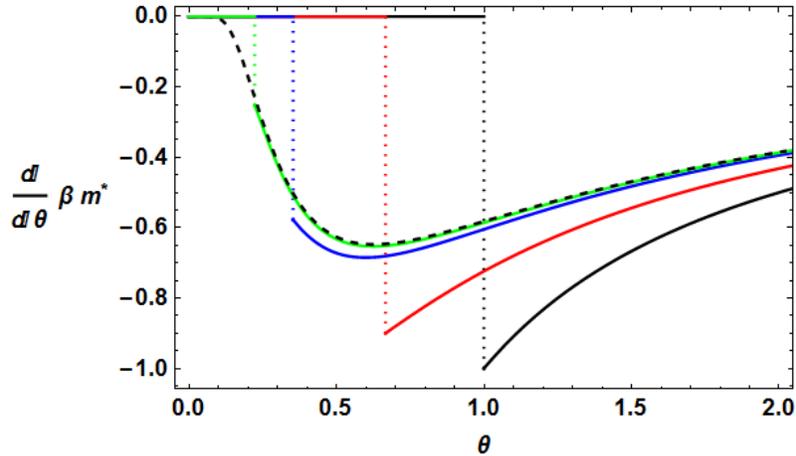

**Figure 5** – Derivative of $\beta m^*$ function with respect to the reduced temperature θ for distinct number of particles *N*, *N* = 2 (black), 3 (red), 10 (blue), 50 (green), and $N \to \infty$ (dashed black line)

# 4 Conclusions

An extended form of the Einstein model for specific heat in solids was proposed taking into account the number of harmonic oscillators (*N*). The system entropy was strictly deduced for arbitrary *N*, enabling the formulation of thermodynamic analogous functions describing the physical properties of finite systems, such as chemical potential and specific heat. The analogous chemical potential, *m*, was obtained using a discrete form of the Leibniz integral rule while the analogous specific heat, $\chi$, by simple differentiation technique.

The critical temperature, where both entropy and chemical potential become null, was obtained and identified as the temperature associated with a Bose-Einstein condensate ($\theta_0$). The internal energy and the chemical potential are non-smoothed functions at $\theta_0$, consequently the first derivative of those energies are discontinuous at $\theta_0$, i.e., a first-order transition like behavior was found. The discontinuities of $\chi(\tau)$ and $\beta\,m^*(\tau)$ derivatives (or non-smoothed behavior of $u(\tau)$ and $\beta\,m^*(\tau)$) appear exclusively for finite number of non-interacting harmonic oscillators (finite size Einstein's solid model) and vanish in the thermodynamic limit. Usually, discontinuities appear only in the thermodynamic limit of a Bose gas, which rigorously disappear for finite systems, e.g., the effect of a finite number of particles in the Bose-Einstein condensation of a trapped gas [13].

For finite size Einstein solids (nano-clusters) the critical temperature is always non-null and its reduced form, $\frac{\tau_0 k_B}{\hbar\omega} = \theta_0$, has an inverse of harmonic series dependence on *N*. The values of $\chi$ in the limit of low-temperatures vary strongly for small values of *N*, but are always above the thermodynamic limit in agreement with experimental results [10]. Our analysis suggests that the simple hypothesis of assuming the distribution of the total internal energy among the oscillators suffices to describe the behavior of nano-systems [10-12]. As a possible direct application of the model proposed here we refer to the specific heat of carbon nanotubes.

# Acknowledgment

The financial support from the São Paulo Research Foundation (FAPESP) grants 2013/25210-8 and 2010/11385-2 is acknowledged. MdS acknowledges support from the National Council of Technological and Scientific Development – CNPq (Grants 305472/2014-3).

# Appendix A – *Simplifying ln Ω*

By applying the logarithm function to both sides of the equation for number of accessible eigen-states ($\Omega$) for generic $M = F(E) = N f(u)$ and $u=E/N$ is

$$\Omega(E,N) = \frac{(F(E)+N-1)!}{(F(E))!(N-1)!}, \tag{A1}$$

one can write:

$$\ln(\Omega(E,N)) = \ln(F(E)+N-1)! - \ln(F(E))! - \ln(N-1)! \tag{A2}$$

or, equivalently:

$$\ln(\Omega(E,N)) = \sum_{i=1}^{F(E)+N-1} \ln(i) - \sum_{j=1}^{F(E)} \ln(j) - \sum_{k=1}^{N-1} \ln(k). \tag{A3}$$

The first two terms on the right side of eq.(A3) can be reduced to one by subtracting the common terms:

$$\ln(\Omega(E,N)) = \sum_{i=F(E)+1}^{F(E)+N-1} \ln(i) - \sum_{k=1}^{N-1} \ln(k). \tag{A4}$$

Finally, by recognizing that the two sums on the right of eq. (A4) have the same number of terms, this equation can be rewritten in the following simplified form:

$$\ln(\Omega(E,N)) = \sum_{k=1}^{N-1} \ln\left(\frac{F(E)+k}{k}\right), \tag{A5}$$

where the number of terms in the sum is energy independent. This is particularly appropriate for the calculations since E is the unknown for given *N* and *τ*. Thus the eq. (A1) can be rewritten as:

$$\Omega(E,N) = \prod_{k=1}^{N-1}\left(\frac{F(E)}{k}+1\right). \tag{A6}$$

The entropy is given by:

$$s=S/N=\frac{k_B}{N}\ln(\Omega(E,N)) = \frac{k_B}{N}\sum_{k=1}^{N-1}\ln\left(\frac{F(E)+k}{k}\right) = \frac{k_B}{N}\sum_{k=1}^{N-1}\ln\left(\frac{f(u)+k/N}{k/N}\right). \tag{A7}$$

and the temperature ($\tau$) is obtained from $\partial s/\partial u = \partial S/\partial E = 1/\tau$ and that is:

$$\frac{1}{\tau} = k_B[\,\partial F(E)/\partial E\,]\sum_{k=1}^{N-1}(F(E)+k)^{-1} = \frac{k_B}{N}[\,\partial f(u)/\partial u\,]\sum_{k=1}^{N-1}\left(f(u)+\frac{k}{N}\right)^{-1}. \tag{A8}$$

Considering $N \to \infty$ and using the Stirling approximation in Eq (A7) and $\partial s/\partial u = \partial S/\partial E = 1/T$ one can obtain the energy Planck distribution for generic $f(u)$ with $\beta = 1/(k_BT)$:

$$f(u) = \frac{1}{exp\left(\frac{\beta}{\partial f(u)/\partial u}\right) - 1}.$$

# Appendix B–*Discrete form of Leibniz integral rule*

When the limits of an integral are not constant, it must be treated according to the Leibniz integral rule, which says that, given a function $\varphi(k, N)$ defined in an integral form:

$$\varphi(k, N) = \int_{a(N)}^{b(N)} f(k, N)\, dk,$$

the derivative of $\varphi(k, N)$ with respect to $N$ is given by:

$$\frac{d\varphi(k,N)}{dN} = \int_{a(N)}^{b(N)} \frac{\partial f(k,N)}{\partial N}\, dk + \frac{db(N)}{dN} f(b(N), N) - \frac{da(N)}{dN} f(a(N), N). \quad (B.1)$$

If $\varphi(k, N)$ is a discrete function of *k* one has:

$$\varphi(k, N) = \sum_{k=a(N)}^{b(N)} f(k, N).$$

Rearranging $\Delta\varphi(k, N) = \varphi(k, N + \Delta N) - \varphi(k, N)$ one obtains:

$$\Delta\varphi(k, N) = \sum_{k=a(N)}^{b(N)} [f(k, N + \Delta N) - f(k, N)] + \sum_{k=b(N)}^{b(N)+\Delta b(N)} f(k, N + \Delta N) + \sum_{k=a(N)+\Delta a(N)}^{a(N)} f(k, N + \Delta N).$$

Finally, the discrete Leibniz sum rule is given by:

$$\frac{\Delta\varphi(k,N)}{\Delta N} = \sum_{k=a(N)}^{b(N)} \frac{[f(k,N+\Delta N) - f(k,N)]}{\Delta N} + \frac{\Delta b(N)}{\Delta N} f(\xi_b, N + \Delta N) - \frac{\Delta a(N)}{\Delta N} f(\xi_a, N + \Delta N), \quad (B.2)$$

where $a(N) \leq \xi_a \leq a(N + \Delta N)$ and $b(N) \leq \xi_b \leq b(N + \Delta N)$.

The discrete derivative in the limit $\Delta N \to 0$ becomes:

$$\frac{d\varphi(k,N)}{dN} = \sum_{k=a(N)}^{b(N)} \frac{df(k,N)}{dN} + \frac{db}{dN} f(b(N), N) - \frac{da}{dN} f(a(N), N).$$
$$(B.3)$$

with *a(N)* = 1, *b(N)* = (*N* - 1), which appear in Eq.(5) as the limits of sum. Hence, the final derivative considered was:

$$\frac{\Delta\varphi(k,N)}{\Delta N} \approx \sum_{k=1}^{N-1} \frac{df(k,N)}{dN} + f(N-1, N). \quad (B.4)$$